\documentclass{article}
\usepackage{amsmath, hyperref, pgfplots}
\usepackage{fullpage}

\DeclareMathOperator{\ud}{d\!}

\begin{document}

\begin{figure}
\begin{flushright}
CTP-SCU/2014001\\
CAS-KITPC/ITP-435
\end{flushright}
\end{figure}

\title{Preferred hierarchy scales from the product landscape}
\author{Songlin Lv\textsuperscript{a, *},
        Zheng Sun\textsuperscript{a, b, \dag} and
        Lina Wu\textsuperscript{b, c, \ddag}\\
        \textsuperscript{a}%
        \normalsize\textit{Center for Theoretical Physics, College of Physical Science and Technology,}\\
        \normalsize\textit{Sichuan University, Chengdu 610064, P. R. China}\\
        \textsuperscript{b}%
        \normalsize\textit{State Key Laboratory of Theoretical Physics and Kavli Institute for Theoretical Physics China (KITPC),}\\
        \normalsize\textit{Institute of Theoretical Physics, Chinese Academy of Sciences, Beijing 100190, P. R. China}\\
        \textsuperscript{c}%
        \normalsize\textit{School of Physical Electronics, University of Electronic Science and Technology of China,}\\
        \normalsize\textit{Chengdu 610054, P. R. China}\\
        \normalsize\textit{E-mail:}
        \textsuperscript{*}\texttt{201322202005@stu.scu.edu.cn,}
        \textsuperscript{\dag}\texttt{sun\_ctp@scu.edu.cn,}
        \textsuperscript{\ddag}\texttt{wulina@std.uestc.edu.cn}
       }
\date{}
\maketitle

\begin{abstract}
The product landscape method has been recently proposed to solve hierarchy problems such as the cosmological constant problem.  We suggest that the parameter distribution on logarithmic scales should be used as a benchmark for hierarchy, and the preferred hierarchy scales can be obtained from the distribution peak.  It is shown that generating hierarchy from purely product distribution is very inefficient.  To achieve a reasonably acceptable efficiency, other effects such as accumulation of weak hierarchy in the effective theory should be incorporated.
\end{abstract}

\section{Introduction}

One of ultimate tasks for theoretical physics is to seek for a fundamental theory which naturally explains all physical phenomena as well as parameters.  Solving hierarchy problems is an essential step towards this task.  Generally speaking, Naturalness of a theory implies that dimensionless free parameters should take values of order $1$, or dimensional parameters should take values close to the fundamental scale of the theory.  So a hierarchy problem occurs when the experimental measurement of a physical parameter is vastly different (usually smaller) than the natural prediction of the theory.  For example, in the hierarchy problem which particle physicists usually refer to~\cite{Wilson:1970ag, Gildener:1976ai, Weinberg:1978ym}, the observed Higgs mass $M_\text{H} \sim 125 \text{GeV}$~\cite{Aad:2012tfa, Chatrchyan:2012ufa} is much lighter than the Planck mass $M_\text{P} \sim 10^{18} \text{GeV}$ or the grand unification scale $M_\text{GUT} \sim 10^{16} \text{GeV}$.  The Higgs mass gets quantum contribution until new physics appears above the scale.  So there is a large discrepancy between the observed Higgs mass and its bare mass, which implies a precise cancellation between the bare mass of the Higgs boson and its quantum correction.  In another well-known example, the cosmological constant problem~\cite{Weinberg:1988cp}, the cosmological constant or the vacuum energy density has its measured value $\Lambda \sim 10^{- 122} M_\text{P}^4$ which is way too smaller than any known mass scales.  How these small numbers could be naturally realized in a fundamental theory remains an open question in both particle physics and cosmology~\cite{Giudice:2008bi, Feng:2013pwa, Giudice:2013yca}.

As we have mentioned in the above examples, hierarchy problems are usually connected to fine-tuning problems by the procedure of canceling two large parameters to get a small quantity.  Since free parameters are not welcome in a fundamental theory, they should be replaced by dynamical fields which are stabilized in the high-energy microscopic theory.  This stabilization happens in the string landscape~\cite{Bousso:2000xa, Grana:2005jc, Douglas:2006es, Blumenhagen:2006ci}, where quantized fluxes on Calabi-Yau manifolds generates a low-energy effective superpotential after compactifying string theory from $10$ to $4$ dimensions.  A huge number of metastable vacua with different low-energy physics are generated by different choices of fluxes, Calabi-Yau manifolds, etc..  One may hope that at least one vacuum has all parameters stabilized at the measured values of our experiments.  It is still unclear that how our world selects the correct vacuum by either the anthropic principle~\cite{Weinberg:1987dv, Agrawal:1997gf} or some dynamical evolution of the universe~\cite{Dine:2007er, Dine:2008jx, Dine:2009tv}.  Nevertheless, we can currently focus on a more well-defined question in the landscape:  Can we get adequate distribution of vacua similar to the real world, so that the selection to the correct one is feasible?

In the string landscape, vacua are usually densely distributed in the region of our interested, and the distribution can be viewed as continuous~\cite{Denef:2004ze, Denef:2004cf}.  If the distribution respecting to parameters is uniform or quite smooth, the possibility for hierarchy to happen is small, since the typical value of a parameter in such a distribution is of order $1$.  This is just a mathematical restatement of the hierarchy problem and naturalness in the framework of vacuum distributions.  And solving hierarchy problems is now rephrased as seeking a mechanism to alter the vacuum distribution so that hierarchy is preferred.  The most commonly known method is to introduce an exponential factor $e^{- a}$ from non-perturbative dynamics, such as in dynamical supersymmetry breaking~\cite{Poppitz:1998vd, Shadmi:1999jy} or Randall-Sundrum models~\cite{Randall:1999ee}.  An order $1$ parameter $a$ can easily give a small scale after the exponential.  Other approaches also exists, including the method of accumulating many copies of weak hierarchy, such as the loop factor $1/(16 \pi^2)$, to a more notable hierarchy.

The product landscape method was recently proposed as a solution to the cosmological constant problem~\cite{Sumitomo:2012wa, Sumitomo:2012vx, Sumitomo:2012cf}.  It is based on Type-IIB flux compactification models where the vacuum energy density $\Lambda$ can be expressed as the product of several parameters.  These parameters, determined by solving metastable vacua from choices of fluxes, usually have smooth distributions covering the origin of the parameter space.  It is observed that their product distribution is singular at the origin.  So a small $\Lambda$ may be preferred and the hierarchy problem of the cosmological constant may be solved in this plot. Though the prediction of $\Lambda$, or its preferred scales in the landscape, remains unknown.

In this work, we suggest that the logarithmic scale of a parameter such as $\Lambda$, rather than the parameter itself, can be used as a benchmark for hierarchy.  The distribution respecting to $\log \Lambda$ can be acquired from the distribution respecting to $\Lambda$ by comparing the number of vacua from the same interval of $\Lambda$ in different coordinate systems.  By this way, the preferred hierarchy scales can be clearly seen from the distribution peak which turns out to be a small but non-zero value.  Notice that this logarithmic plot criteria applies not only to the product landscape discussed here, but also to a wide range of vacuum distributions from other mechanisms.

From the logarithmic distribution plot, it can be shown that the pure effect of product distributions, excluding the factor from accumulating weak hierarchy introduced in the low-energy effective theory, is quite inefficient in terms of the needed number of variables in the product.  To achieve a reasonably acceptable efficiency, the product landscape method should incorporate other effects which introduce some initial weak hierarchy.  Such weak hierarchy is naturally present between parameters and the fundamental cutoff scale in any effective theory, and can be accumulated in the product form of physical quantities to solve the hierarchy problem.  Notice that our argument is model-independent, so applies not only to the cosmological constant problem based on flux compactification, but also to a wide range of models where the product landscape can be generated in various ways.

The rest content of the paper is organized as following.  Section 2 reviews the product landscape method in previous literature, and its main result for the cosmological constant problem.  Section 3 investigates the logarithmic distribution of the product landscape, calculates the preferred hierarchy scales as the the number of parameters in the product varies, and extracts the pure effect and efficiency of this method excluding other factors.  Section 4 discusses other effects happening in the low-energy effective theory, argues that the accumulation of weak hierarchy plays the major role to solve the hierarchy problem.

\section{Peaking from the product landscape}

The conception of the product landscape method is based on the product distribution in probability theory~\cite{Feller:1968book}, i.e., the probability distribution of the product of several random variables.  Consider a set of random variables $\{x_i | i = 1, \dotsc, n\}$ with probability distributions $P(x_i)$.  The distribution of their product $z = x_1 \dotsm x_n$ can be calculated as
\begin{equation}
\begin{split}
P(z) &= \int P(x_1) \dotsm P(x_n) \delta(z - x_1 \dotsm x_n) \ud x_1 \dotsm \ud x_n\\
     &= \int P(x_1) \dotsm P(x_{n - 1}) \frac{P(z / (x_1 \dotsm x_{n - 1}))}{x_1 \dotsm x_{n - 1}} \ud x_1 \dotsm \ud x_{n - 1} .
\end{split}
\end{equation}
For simplicity, we take the distributions of $x_i$'s to be uniform on $(0, 1)$, i.e.,
\begin{equation} \label{eq:01}
P(x_i) =
\begin{cases}
1, &\text{for $x_i \in (0, 1)$} ,\\
0, &\text{for $x_i \le 0$ or $x_i \ge 1$} .
\end{cases}
\end{equation}
This leads to the product distribution
\begin{equation}
P(z) =
\begin{cases}
(- \log z)^{n - 1} / (n - 1)!, &\text{for $z \in (0, 1)$} ,\\
0,                             &\text{for $z \le 0$ or $z \ge 1$} .
\end{cases}
\end{equation}

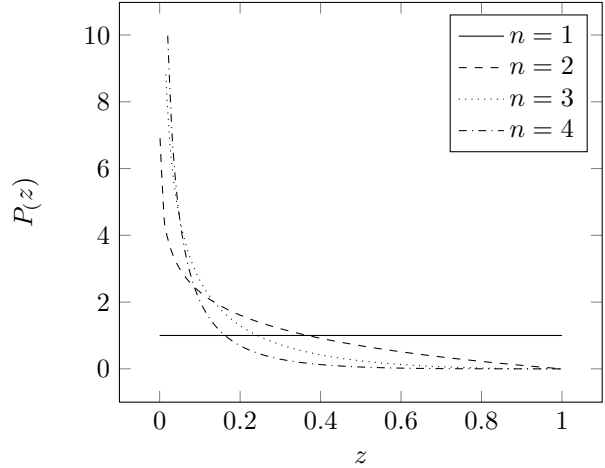
\begin{figure}
\centering
\begin{tikzpicture}
\begin{axis}[xlabel = $z$, ylabel = $P_(z)$, width = 8cm, legend pos = north east]
\addplot[domain = 0 : 1]{1};
\addplot[domain = 1e-3 : 1, samples = 100, smooth, style = {dashed}]{- ln(x)};
\addplot[domain = 1.5e-2 : 1, samples = 100, smooth, style = {dotted}]{(- ln(x))^2 / 2};
\addplot[domain = 2e-2 : 1, samples = 100, smooth, style = {dashdotted}]{(- ln(x))^3 / 6};
\legend{$n = 1$, $n = 2$, $n = 3$, $n = 4$};
\end{axis}
\end{tikzpicture}
\caption{The distribution $P(z)$, where $z = x_1 \dotsm x_n$ and $x_i$'s have uniform distributions on $(0, 1)$.} \label{fg:01}
\end{figure}

The distribution $P(z)$ has a singular peak at the origin, which implies that small values of $z$ may be preferred.  As $n$ goes larger, the distribution becomes more singular and a small $z$ seems to be more preferred.  Such behavior of $P(z)$ can be seen in the plot of Figure \ref{fg:01}.

\begin{table}
\centering
\begin{tabular}{|c|c|}
\hline
$z$                  & $P(z)$ near $z = 0$\\
\hline
$x_1 \dotsm x_n$     & $\displaystyle \frac{1}{(n - 1)!} (- \log z)^{n - 1}$\\
$x_1^n$              & $\displaystyle \frac{1}{n} z^{- 1 + 1 / n}$\\
$x_1^n \dotsm x_m^n$ & $\displaystyle \frac{1}{n^m (m - 1)!} z^{ -1 + 1 / n} (- \log z)^{m - 1}$\\
$x_1^n x_2^m$        & $\displaystyle \frac{1}{n - m} (z^{- 1 + 1 / n} - z^{- 1 + 1 / m})$\\
$x_1^n / x_2^m$      & $\displaystyle \frac{1}{n + m} z^{- 1 + 1 / n}$\\
$x_1^n + x_2^m$      & $\displaystyle \frac{\Gamma(1 + 1 / n) \Gamma(1 + 1 / m)}{\Gamma(1 / n + 1  / m)} z^{- 1 + 1 / n + 1 / m}$\\
\hline
\end{tabular}
\caption{Behavior of distributions $P(z)$ near the origin, with different expressions of $z$.  All $x_i$'s have uniform distributions on $(0, 1)$.} \label{tb:01}
\end{table}

In general, distributions of quantities involving products and powers of $x_i$'s always have such a singularity and become more singular as the number of $x_i$'s goes larger.  Several other examples are listed in Table \ref{tb:01}.  Notice also that sums in the expression of $z$ make the distribution less singular than the distribution of each term, thus should be avoided when trying to solve hierarchy problems.

Although our calculation is based on uniform distributions of $x_i$'s on $(0, 1)$, the main result, that the product distribution has a singular peak at the origin, can be extended to cases where distributions of random parameters are smooth near the origin.  Moreover, if some of $x_i$'s already have singular distributions at the origin, the singularity in the product distribution will be fortified and hierarchy will be statistically more preferred.

These arguments can be realized in Type-IIB flux compactification models, where the vacuum energy density $\Lambda$ is expressed as the product of several parameters calculated from the low-energy effective supergravity theory.  Parameters are generated from either discrete but densely distributed fluxes, or non-perturbative dynamics~\cite{Kachru:2003aw, Balasubramanian:2005zx}.  In most cases, their distributions cover the origin, and are either smooth or singular at the origin.  So from the arguments above, a small $\Lambda$ seems being preferred in the landscape of vacua.  In some models with a large number of complex structure moduli, the expectation values of $\Lambda$ can be comparable with the observed cosmological constant \cite{Sumitomo:2012vx}.  Thus one may hopefully expect that a promising prediction from the string theory could be made through this product landscape procedure.

\section{Preferred hierarchy on logarithmic scales}

Although the singular behavior of the product distribution indicates some preference of hierarchy, it does not give a specific prediction of the hierarchy scale.  Works in previous literature~\cite{Sumitomo:2012wa, Sumitomo:2012vx, Sumitomo:2012cf, Sumitomo:2013vla} suggest either the expectation value $\langle z \rangle$ or the value $z_{Y \%}$ with cumulative probability $Y \%$ as a measure of the peaky behavior, and calculate their dependence on $n$ as a rating of the method.  But the prediction question is still not answered without invoking some degree of anthropic selection.  One may naively think that the occurrence of the peak, at the origin or some low cutoff of the effective theory, gives the preference in the distribution.  But as we are to show now, when reading the preference information, one should be careful about the coordinate system on which the distribution is calculated.

When speaking of hierarchy, it is the scale of a parameter instead of the parameter value itself that reflects how strong the hierarchy is.  For example, if we are interested in some phenomenon with hierarchy of $10^{- 10}$, what we expect from the theoretical explanation is that the possibility for some parameter falling into the scale region around our interested one, such as $(10^{- 11},10^{- 9})$, would be higher than the possibility in other regions, such as $(10^{- 3}, 10^{- 1})$.  Although these regions span scale intervals of the same size, their size on a linear coordinate system is not comparable.  To properly compare the distribution, a logarithmic coordinate system is needed, so that these scale intervals of our interest are kept the same size.  Hence the distribution $P(z)$ should be reinterpreted on logarithmic scales as $P(\log z)$, or noted as $P_{\log}(z)$.

Consider a small region of size $\ud z$ near $z$.  The probability or number of vacua in this region should be the same when viewing from different coordinate systems.  Comparing linear and logarithmic coordinates, we have
\begin{equation}
P(z) \ud z = P_{\log}(z) \ud \, \log z .
\end{equation}
This gives the logarithmic distribution
\begin{equation}
P_{\log}(z) = z P(z) .
\end{equation}
We can already see an important consequence from this expression:  The linearly uniform distribution \eqref{eq:01} actually prefers the highest logarithmic scale at $\log z = 0$, or $z = 1$.  The distribution $P(z) = z^{- 1}$, which is singular at the origin, is the actual uniform distribution $P_{\log}(z) = 1$ on all scales \footnote{Notice that $P(z) = z^{- 1}$ is not normalizable without introducing cutoffs.  So in reality we have a uniform distribution in a scale interval.}, as mentioned in the intermediate scale branch of the landscape~\cite{Dine:2004is, Dine:2005yq, Dine:2005iw}.  One may have noticed previously that all distributions calculated in Table \ref{tb:01} are less singular than $z^{- 1}$.  So no matter how many random parameters and their powers are multiplied together, the resulting distribution always has a preferred scale which is at neither the origin nor the low cutoff.

Now the preferred hierarchy scale $z_0$ can be readily identified by the peak of the distribution on logarithmic scales.  This can be done by solving the stationary condition at the peak
\begin{equation} \label{eq:02}
\frac{\partial}{\partial z} P_{\log}(z_0) = 0 .
\end{equation}
Strictly speaking, one should check the second derivative to ensure the stationary point to be a maximum.  In some cases, the end points of the distribution range should also be checked.  But in most cases when the peak can be seen and estimated from the distribution plot, checking the stationary condition \eqref{eq:02} would be adequate.

Before proceeding, we would like to emphasize that the above criteria applies not only to the product landscape discussed here, but also to a wide range of vacuum distributions from other mechanisms.  When the scale of some physical quantity rather than its value is our concern, the distribution should be plotted on a logarithmic coordinate system so that the peak can be properly identified and expressed as the preference of the distribution.  This argument has been widely accepted in the study of string landscape and actually contributes to the three branches of the landscape~\cite{Dine:2005yq}.

Let us return to the distributions in Table \ref{tb:01}.  As an example, for the product distribution of $z = x_1 \dotsm x_n$ with uniform distributions of $x_i$'s on $(0, 1)$, we have
\begin{equation} \label{eq:03}
P_{\log}(z) = z P(z) =
\begin{cases}
z (- \log z)^{n - 1} / (n - 1)!, &\text{for $z \in (0, 1)$} ,\\
0,                               &\text{for $z \le 0$ or $z \ge 1$} .
\end{cases}
\end{equation}
Solving the stationary condition \eqref{eq:02} gives the peak at
\begin{equation}
z_0 = e^{1 - n} .
\end{equation}
The position of the peak becomes exponentially smaller as $n$ goes larger.  So it seems quite plausible to solve the hierarchy problem in this framework.

However, there is another factor affecting the occurrence of the peak.  The assumption that $x_i$'s have uniform distributions on $(0, 1)$ gives an expectation value $\langle x_i \rangle = 1 / 2$.  Thus we already have some weak hierarchy from the start, and the product procedure is accumulating this factor.  For $z = x_1 \dotsm x_n$ we have $\langle z \rangle = 2^{- n}$ which goes exponentially small as n goes larger.  To exclude this factor and see the pure effect of the product distribution, $x_i$'s should be assumed to have uniform distributions on $(0, 2)$, which gives $\langle x_i \rangle = 1$.  Then the distribution \eqref{eq:03} is modified to
\begin{equation} \label{eq:04}
P_{\log}(z') =
\begin{cases}
2^{- n} z' (- \log (2^{- n} z'))^{n - 1} / (n - 1)!, &\text{for $z' \in (0, 2^n)$} ,\\
0,                                                   &\text{for $z' \le 0$ or $z' \ge 2^n$} .
\end{cases}
\end{equation}
Although the modified product $z'$ satisfies $\langle z' \rangle = 1$, its distribution \eqref{eq:04} leads to the peak at
\begin{equation} \label{eq:05}
z'_0 = 2^n e^{1 - n}
\end{equation}
which becomes exponentially smaller as $n$ goes larger.  This is the actual hierarchy which we can acquire purely from the product distribution.

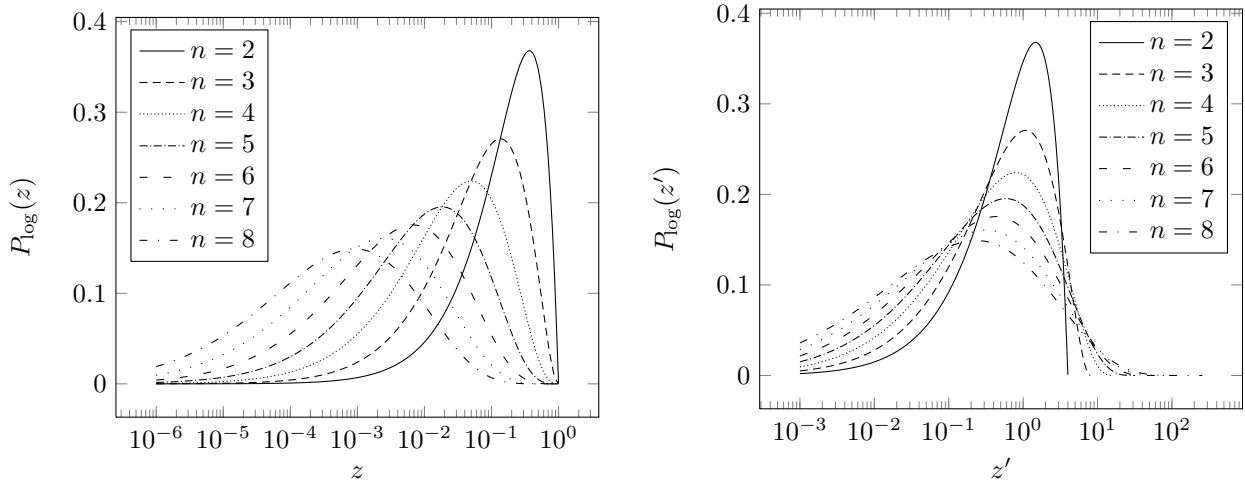
\begin{figure}
\centering
\begin{tikzpicture}
\begin{semilogxaxis}[xlabel = $z$, ylabel = $P_{\log}(z)$, width = 8cm, legend pos = north west]
\addplot[domain = 1e-6 : 1, samples = 100, smooth]{- x * ln(x)};
\addplot[domain = 1e-6 : 1, samples = 100, smooth, style = {densely dashed}]{x * (- ln(x))^2 / 2};
\addplot[domain = 1e-6 : 1, samples = 100, smooth, style = {densely dotted}]{x * (- ln(x))^3 / 6};
\addplot[domain = 1e-6 : 1, samples = 100, smooth, style = {densely dashdotted}]{x * (- ln(x))^4 / 24};
\addplot[domain = 1e-6 : 1, samples = 100, smooth, style = {loosely dashed}]{x * (- ln(x))^5 / 120};
\addplot[domain = 1e-6 : 1, samples = 100, smooth, style = {loosely dotted}]{x * (- ln(x))^6 / 720};
\addplot[domain = 1e-6 : 1, samples = 100, smooth, style = {loosely dashdotted}]{x * (- ln(x))^7 / 5040};
\legend{$n = 2$, $n = 3$, $n = 4$, $n = 5$, $n = 6$, $n = 7$, $n = 8$};
\end{semilogxaxis}
\end{tikzpicture}
\hfill
\begin{tikzpicture}
\begin{semilogxaxis}[xlabel = $z'$, ylabel = $P_{\log}(z')$,  width= 8cm, legend pos = north east]
\addplot[domain = 1e-3 : 4, samples = 100, smooth]{- x * ln(x / 4) / 4};
\addplot[domain = 1e-3 : 8, samples = 100, smooth, style = {densely dashed}]{x * (- ln(x / 8))^2 / 2 / 8};
\addplot[domain = 1e-3 : 16, samples = 100, smooth, style = {densely dotted}]{x * (- ln(x / 16))^3 / 6 / 16};
\addplot[domain = 1e-3 : 32, samples = 100, smooth, style = {densely dashdotted}]{x * (- ln(x / 32))^4 / 24 / 32};
\addplot[domain = 1e-3 : 64, samples = 100, smooth, style = {loosely dashed}]{x * (- ln(x / 64))^5 / 120 / 64};
\addplot[domain = 1e-3 : 128, samples = 100, smooth, style = {loosely dotted}]{x * (- ln(x / 128))^6 / 720 / 128};
\addplot[domain = 1e-3 : 256, samples = 100, smooth, style = {loosely dashdotted}]{x * (- ln(x / 256))^7 / 5040 / 256};
\legend{$n = 2$, $n = 3$, $n = 4$, $n = 5$, $n = 6$, $n = 7$, $n = 8$};
\end{semilogxaxis}
\end{tikzpicture}
\caption{Left:  The distribution $P(z)$ on logarithmic scales, i.e., $P_{\log}(z)$, where $z = x_1 \dotsm x_n$ and $x_i$'s have linearly uniform distributions on $(0, 1)$.  Right:  The modified distribution $P_{\log}(z')$, where $z' = x_1 \dotsm x_n$ and $x_i$'s have linearly uniform distributions on $(0, 2)$, thus $\langle x_i \rangle = 1$ and $\langle z' \rangle = 1$.} \label{fg:02}
\end{figure}

The distributions \eqref{eq:03} and \eqref{eq:04} are plotted in Figure \ref{fg:02}.  It can be seen that although the peak moves towards smaller scales as $n$ increases, the movement becomes very slow after the accumulation factor $\langle z \rangle = 2^{-n}$ is excluded.  Numerically \eqref{eq:05} shows that $n \sim 10$ variables are needed to be multiplied together to get a hierarchy of $z'_0 \sim 0.1$.  If hierarchy is generated exclusively by this means, the cosmological problem $\Lambda / M_\text{P}^4 \sim 10^{-122}$ needs $n \sim 1000$, and the hierarchy problem of Higgs mass $M_\text{H} / M_\text{GUT} \sim 10^{-14}$ needs $n \sim 100$.  Such huge number of variables complicate the theory, and it is quite uncommon for a model to express a physical quantity as a single term of product of many factors.  So one is actually sacrificing much simplicity of the model for little hierarchy, making the theory unnatural from another perspective~\cite{ArkaniHamed:2012gw}.  In summary, our analysis shows that the product landscape method is very inefficient to solve the hierarchy problem if no other mechanism is involved.

\section{Accumulation effects in effective theories}

As the previous section has shown, to achieve a reasonably acceptable efficiency, the product landscape method should incorporate other effects which introduce some initial weak hierarchy.  As the number of parameters $n$ in the product increases, accumulation of such small effects generates hierarchy exponentially, which overwhelms the effect from product distributions.  This is what actually happens in many low-energy effective theories.  For example, in some Type-IIB flux compactification models, the cosmological constant $\Lambda$ can be expressed as the product of many coefficients of the superpotential which is generated from fluxes.  To keep the low-energy effective superpotential description valid, magnitudes of coefficients in the superpotential should not exceed the string scale.  So we naturally have a weak hierarchy below the string scale and it can be accumulated in $\Lambda$.  If we assume the initial weak hierarchy scale is $1/2$, the hierarchy problem of the cosmological constant $\Lambda / M_\text{P}^4 \sim 10^{-122}$ requires $n \sim 400$, which is comparable with the result from multi-moduli cases of Type-IIB flux compactification models, where $n$ corresponds to the number of complex structure moduli~\cite{Sumitomo:2012vx}.  The effect from product distribution contributes to the hierarchy as well.  If the effect of multiple moduli stabilization~\cite{Shiu:2011zt, Rummel:2013yta} in the complex structure sector is also included, the required number of moduli can be reduced to $n \sim 200$.  And such number of complex structure moduli is present in many constructions of Calabi-Yau manifolds by complete intersections~\cite{Hubsch:1992nu, cy:url}.

The above argument for Type-IIB flux compactification models can be extended to more general effective theories.  Any effective theory has a scale above which new physics appears.  Parameters in the theory should be kept below such scale to keep the effective description valid.  Then there is already a weak hierarchy between parameters and the fundamental cutoff scale of the theory.  With proper model building, the physical quantity which we are interested in may be expressed as a product form, and the initial weak hierarchy may be accumulated to solve the hierarchy problem.  One may also consider introducing the initial weak hierarchy by other means, and explore various types of hierarchy problems through the accumulation effect combined with the product landscape method.

\section*{Acknowledgement}

We thank Tianjun Li, Muyang Liu, Yoske Sumitomo and Henry Tye for helpful discussions.  This work is supported by the National Natural Science Foundation of China under grant 11305110.

\end{document}